# AN IRON-RICH SUN AND ITS SOURCE OF ENERGY*

O. Manuel[‡] and A. Katragada

*Departments of Chemistry and Computer Engineering, University of Missouri, Rolla, MO 65401, USA*

**Abstract:** Mass-fractionation enriches light elements and the lighter isotopes of each element at the solar surface, making a photosphere that is 91% H and 9% He. However, the solar interior consists mostly of elements that comprise 99% of ordinary meteorites – Fe, O, Ni, Si, S, Mg and Ca – elements made in the deep interior of a supernova. Solar energy arises from a series of nuclear reactions triggered by neutron-emission from the collapsed supernova core on which the Sun formed. Solar mass-fractionation, solar neutrinos, and an annual outpouring of $3 \times 10^{43}$ H atoms in the solar wind are by-products of solar luminosity.

## 1. INTRODUCTION

*"The sun is the Rosetta stone of astrophysics,"* says Göran Scharmer, director of Sweden's Institute for Solar Physics, in a news report for the July 2004 issue of <u>National Geographic Magazine</u> [1]. *"But it is a stone that we haven't been able to decrypt entirely."*

Aston [2] provided the key to this puzzle in 1913. That year he noted that the atomic weight of Neon is lighter after diffusing through pipe clay walls, as if Neon had isotopes of different mass. Fifty-six years later, Neon of light atomic weight was found in the surfaces of Moon samples returned by the Apollo mission [3]. Subsequent measurements [4], shown in Figure 1, revealed that lighter mass (L) isotopes of all noble gases in the solar wind are enriched relative to the heavier (H) ones by a common fractionation factor ($f$), where

$$f = (H/L)^{4.56} \qquad (1)$$

When this fractionation power law is applied to the abundance pattern of elements in the photosphere [5], the most abundant elements in the interior of the Sun turn out to be Fe, O, Ni, Si, S, Mg and Ca (Figure 2) - elements that are also most abundant in planets close to the Sun.

They are the same, even-numbered elements Harkins found in 1917 [6] to comprise 99% of ordinary meteorites. <u>*The probability (P) of fortuitous agreement is $P < 2 \times 10^{-33}$*</u> [7].

The iron-rich Sun explains solar eruptions and magnetic fields [8] and answers the question, *"What kind of environment gave birth to the Sun and planets?"*, raised by live $^{60}$Fe in the early solar system from a supernova (SN) core. A supernova exploded here 5 billion years ago [10]. The Sun formed on the collapsed SN core. Iron meteorites, cores of terrestrial planets, and the interior of the Sun consists mostly of Fe and other elements made near the SN core [11].

*Work supported by the Foundation for Chemical Research and the University of Missouri-Rolla
[‡]*E-mail:* om@umr.

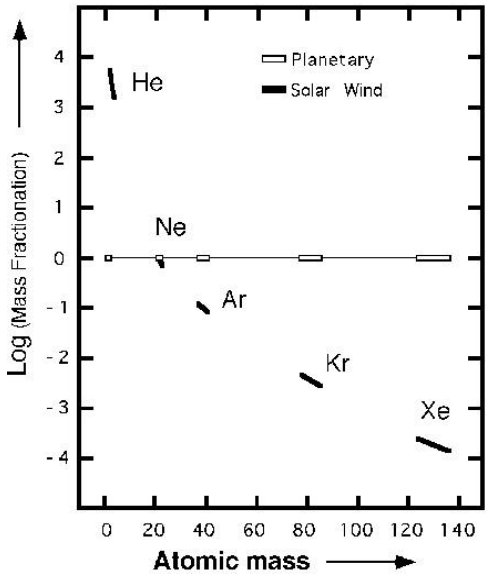 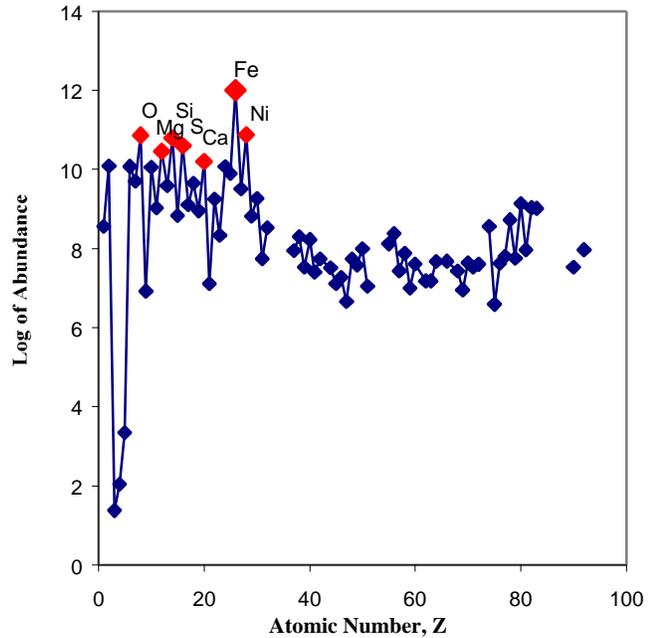

Figure 1. Light isotopes of all noble gases are enriched in the solar wind [4].

Figure 2. Internal composition of the Sun [4] after correcting for mass fractionation seen across isotopes.

## 2. SOURCE OF LUMINOSITY

Nucleons in Fe, Ni, O, Si, S, Mg and Ca are too tightly bound to generate solar luminosity. However systematic properties of the 2,850 known nuclides [12] (Figure 3), when considered in terms of reduced variables like Z/A (charge per nucleon) and M/A (mass or potential energy per nucleon), reveal an inherent instability of the collapsed SN core toward neutron-emission [13-16]. This process releases 10-22 MeV per neutron emitted [16], which then triggers a series of reactions that collectively produce solar luminosity (SL), solar neutrinos, an upward flow of $H^+$ "carrier" ions that maintains mass separation in the Sun, and then departs in the solar wind (SW):

a) Escape of neutrons from the collapsed solar core
$$<_0^1n> \rightarrow {}_0^1n + \sim 10\text{-}22 \text{ MeV} \quad (>57\% \text{ SL})$$

b) Neutron decay or capture by other nuclides
$$_0^1n \rightarrow {}_1^1H^+ + e^- + \text{anti-}\nu + 0.78 \text{ MeV} \quad (<5\% \text{ SL})$$

c) Fusion and upward acceleration of $H^+$ by deep-seated magnetic fields
$$4 \; _1^1H^+ + 2 \; e^- \rightarrow {}_2^4He^{++} + 2\nu + 27 \text{ MeV} \quad (<38\% \text{ SL})$$

d) Escape of excess $H^+$ that survives the upward journey in the solar wind
$$3 \times 10^{43} \; H^+/\text{year depart in the solar wind} \quad (100\% \text{ SW})$$

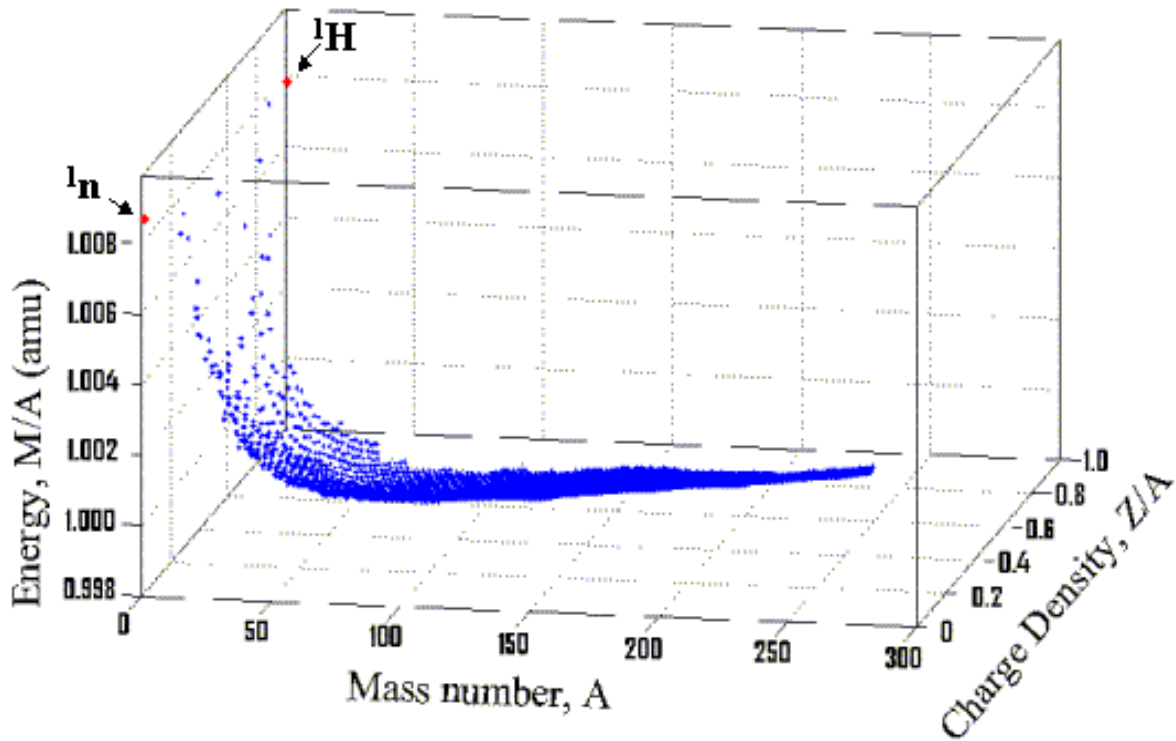

Figure 3. Systematic properties of the 2,850 known nuclides reveal neutron repulsion that drives neutron-emission from the collapsed SN-core on which the Sun formed [15].

In addition to explaining solar luminosity (SL), solar neutrinos, an upward flow of $H^+$ "carrier" ions that maintains mass separation in the Sun and then departs in the solar wind (SW) [13-16], recent observations suggest that steps a) – d) occur in the Sun and other Sun-like stars:

1. The *number of solar electron neutrinos measured is <38%* of the number expected if step (c), H-fusion alone, produced all solar luminosity [16,17].

2. A recent survey [18] of other Sun-like stars during the "Maunder minimum", a period of low sun-spot and magnetic storm activity, found that the *stars appeared to be metal-rich, as expected if mass separation depends on magnetic fields for upward acceleration of the carrier gas* ($H^+$ *ions*) *from the interior of the star* [7].

Regarding observation #1, the electron neutrinos measured in the Charged Current reaction [16,17] at the Sudbury Neutrino Observatory (SNO) represent >87% of the electron neutrinos produced by H-fusion in the Sun if processes a) - d) occur there. If the standard solar model were correct, neutrino oscillations produce the neutral current observed in the SNO experiment.

Regarding observation #2, the "Maunder minimum" refers to the 70-year period from 1645 to 1714 when there was very little sunspot activity during the coldest part of the Little Ice Age in Europe and North America. Wright notes in a news report on their survey of other stars that *"the vast majority of stars identified as Maunder minimum stars . . . are either evolved stars or stars rich in metals like iron and nickel."*

## 3. CONCLUSIONS

Mass-fractionation enriches light elements and the lighter isotopes of each element at the solar surface, making a photosphere that is 91% H and 9% He. However, the solar interior consists mostly of seven, even-numbered elements of high nuclear stability - Fe, O, Ni, Si, S, Mg and Ca. These elements were made in the deep interior of the supernova that gave birth to the solar system 5 billion years ago. They comprise 99% of ordinary meteorites.

Solar energy arises from a series of nuclear reactions triggered by neutron-emission from the collapsed supernova core on which the Sun formed. Solar mass-fractionation, solar neutrinos, and an annual outpouring of $3 \times 10^{43}$ H atoms in the solar wind are by-products of solar luminosity.

## REFERENCES

1. C. Suplee, News release on feature story in *National Geographic Magazine* (July 2004) http://magma.nationalgeographic.com/ngm/0407/feature1/index.html
2. F. W. Aston, *British Assoc. Adv. Sci. Reports* **82** (1913) 403.
3. LPET, *Science* **165** (1969) 1211-1227.
4. O. Manuel and G. Hwaung, *Meteoritics* **18** (1983) 209-222.
5. J. E. Ross and L. H. Aller, *Science* **191** (1976) 1223-1229.
6. W. D. Harkins, *J. Am. Chem. Soc.* **39** (1917) 856-879.
7. O. Manuel and S. Friberg, *ESA SP-517*: *Proc. 2002 SOHO/GONG. Conf.* (2003) 345-347.
8. O. Manuel, B.W. Barry and S.E. Friberg, *J. Fusion Energy* **21** (2002) 193-198.
9. J.J. Hester et al., *Science* **304** (2004) 1116-1117.
10. P.K. Kuroda and W.A. Myers, *Naturwissenschaften* **85** (1998) 180.
11. O.K. Manuel and D.D. Sabu, *Trans. Mo. Acad. Sci.* **9** (1975) 104; *Science* **195** (1977) 208.
12. J. K. Tuli, *Nuclear Data Cards,* 6$^{th}$ ed. (Brookhaven Nat'l Lab, 2000) 74 pp.
13. O. Manuel et al., *J. Fusion Energy* **19** (2000) 93-98.
14. O. Manuel et al., *32$^{nd}$ Lunar Planet Sci. Conf.* (2001) Abstract #1041.
15. O. Manuel, E. Miller and A. Katragada, *J. Fusion Energy* **20** (2001) 197-201.
16. O. Manuel, C. Bolon and M. Zhong, *J. Radioanal. Nucl. Chem.* **252** (2002) 3-7.
17. Q. R. Ahmad et al., *Phys. Rev. Lett.* **89** (2002) 011301.
18. M. B. Coelin (ed.), *Proc. 2nd NO-VE Workshop on Neutrino Oscillations* (Venice, 2003) 553 pp + 4 page addendum.
19. J. Wright and G. Marcy, news report on paper presented at the 2004 AAS Meeting in Denver (31 May 2004) http://www.berkeley.edu/news/media/releases/2004/06/01_maunder.shtml